\documentclass[a4paper]{llncs}

\usepackage{amssymb}
\usepackage{multirow}
\usepackage{graphicx}
\usepackage[font=scriptsize,labelfont=bf]{caption}
\usepackage{amsmath,amssymb,amsfonts, subfigure}
\usepackage{comment}
\usepackage{algorithmicx}
\usepackage[ruled]{algorithm}
\usepackage{algpseudocode}
\renewcommand{\algorithmicrequire}{\textbf{Input:}}
\renewcommand{\algorithmicensure}{\textbf{Output:}}

\begin{document}

\title{Modeling Memetics using Edge Diversity}
\author{Yayati Gupta\inst{1}, Akrati Saxena\inst{1}, Debarati Das\inst{2}
\and S. R. S. Iyengar\inst{1}}
\institute{Department of Computer Science, Indian Institute of Technology Ropar\\
\email{yayati.gupta@iitrpr.ac.in, akrati.saxena@iitrpr.ac.in, sudarshan@iitrpr.ac.in}
\and
Department of Computer Science, PES Institute of Technology, Bangalore\\
\email{debarati.d1994@gmail.com }}

\maketitle              

\begin{abstract}

The study of meme propagation and the prediction of meme trajectory are emerging areas of interest in the field of complex networks research. In addition to the properties of the meme itself, the structural properties of the underlying network decides the speed and the trajectory of the propagating meme. In this paper, we provide an artificial framework for studying the meme propagation patterns. Firstly, the framework includes a synthetic network which simulates a real world network and acts as a testbed for meme simulation. Secondly, we propose a meme spreading model based on the diversity of edges in the network. Through the experiments conducted, we show that the generated synthetic network combined with the proposed spreading model is able to simulate a real world meme spread. Our proposed model is validated by the propagation of the Higgs boson meme on Twitter as well as many real world social networks. 
\end{abstract}

\section{Introduction}

``We ape, we mimic, we mock, we act'' is a law universal to all human beings. Imagine a lady in an elevator, heading to the fifth floor of her office. Suddenly, one by one, every person in the elevator turns back, what does she do now? According to Elevator Groupthink psycology experiment \cite{clissold2004candid}, most of us would turn back in such a situation. Usually, most of us become followers of the crowd when faced with our sense of conformity. If ants follow each other with the help of the pheromone trail, humans too involuntarily imitate and follow each other’s' behaviours and ideas. Behaviours like obesity, smoking and altruism are also seen to spread through social networks \cite{christakis2007spread}. Today, Online Social Networks(OSNs) like Facebook and Twitter provide a platform to fulfill people's penchant for information sharing, arguing and mudslinging. Used by approximately 1.4 billion people worldwide \cite{statisticsstatistic}, Facebook's ``Read, Like and Share'' tradition has today become a way of living. Understanding these spreading phenomena can help us in diverse ways such as accelerating the spread of useful information i.e. health related advices or disaster management related announcements as well as for viral marketing of products and memes. Predicting the trajectory of a meme's propagation in a network can also prevent the spread of malicious rumors and misinformation. Social networks play an instrumental role in the spread of influence in today's world. Hence, contagion prediction models are an extensively studied field in complex networks research. Such models evolve frequently with time, aiming to depict real world information propagation more accurately. Initially, meme propagation models were inspired from compartmental epidemiological models \cite{daley1964epidemics}. These models \cite{hethcote1989three} were too simplistic and did not consider the role of edges in the spreading of information. Later on, the advent of independent cascade \cite{goldenberg2001talk} and linear threshold models  \cite{granovetter1978threshold} proved seminal and these became the standardised models for meme propagation. However, most of these models did not take into consideration the network structure and the calculation of parameters for these models also remained a challenge. 

Consider an anecdote about a small child Bob who went to visit the theme park, Six Flags Magic Mountain in California, with his parents. Bob got lost in the “Fright Fest”, which is the biggest and most terrifying maze at the theme park known for its complex spider-web like structure. Confused by the many turns the maze took at every step, poor Bob could not find his way out of the Fright Fest. When Bob did not return, his worried parents contacted the park authorities for help. These authorities having complete knowledge about the structure of the maze and the possible paths that could be traversed by the players, could easily locate Bob. Similarly, real world networks also have a complex yet distinct structure and if one could understand this structure and estimate the paths that can be taken by the meme in its trajectory, could she also not behave like the park authorities in the above analogy?
In connection with the above anecdote, the knowledge of a network's structure is important for understanding meme propagation. It is known that the real world social networks have a very well defined structure. We employ this well known structure for the simulation of a meme. \\
The major contributions of the paper are :
\begin{enumerate}
\item Generation of an artificial synthetic network that mimics a real world social network in terms of network structure. 
\item We propose a spreading model for meme propagation based on the structure of the network. This model is based on the difference in spreading probabilities of different edges which is recognised from the network itself.  
\end{enumerate}
The proposed synthetic network and spreading model give a synthetic simulation environment which serves as a test-bed to study meme propagation patterns. Further, it gives a way of organising the edges in an hierarchy based on varying probabilities of information transmission across these edges.  We validate the proposed spreading model against the real world spreading of the Higgs boson meme on Twitter. If one could extract the structure of offline social networks, our framework can be used for understanding a wide range of phenomena on offline networks as well in addition to online networks. In addition to controlling information flow on OSNs, we can decrease the increasing behavioral spreading of obesity and depression in the world and promote altruism and positive movements. Inspired from the diversity of edges in a social network, the paper lays light on a novel aspect of looking at information propagation.

The rest of the paper is organised as follows: Section 2 describes the related work. Section 3 explains the synthetic networks in addition to describing the real world networks used for simulation. The network structure based spreading model is proposed in Section 4. Section 5 is devoted to results and discussion. Section 6 illustrates the extension of our model as well as future possibilities. Finally, the paper is concluded in Section 7.

\section{Related Work}
An enormous amount of work has been done to study the information propagation pattern on an online social network \cite{bakshy2012role,digg}.  Initially, memes in a social network were considered analogous to a virus in a biological network \cite{daley1964epidemics}. As a result, most information spreading models were inspired from compartmental epidemiological models like SIS and SIR models \cite{hethcote1989three} introduced in 1989. However these models assumed a homogenous mixing of people constituting the population and did not take into account interactions between the individuals. Later, independent cascade(IC) \cite{goldenberg2001talk} and linear threshold(LT) \cite{granovetter1978threshold} models were investigated which are now used as the standard models for information propagation \cite{kleinberg2007cascading}. However, these models did not consider factors like network structure and model simulation parameters.
There were some studies that predicted the parameters associated with the information propagation models \cite{saito2008prediction}, but these are largely based on the utilisation of the past data, obtaining which is a difficult process.  Studying considering the impact of network structure on a meme's propagation provide a relative view of the meme spread. For example, the spread of epidemics is faster on scale free networks as compared to the random networks due to the presence of hubs \cite{pastor2001epidemic}. Zhang et al. presents a stochastic model for the information propagation phenomenon \cite{zhang2011research}. Studying the information propagation may help the scientists in a number of ways like halting the spread of misinformation \cite{budak2011limiting} and accelerating useful information \cite{scanfeld2010dissemination} through a network. 

Meme Virality prediction is an active research area in social network analysis \cite{textviral,natures} and meme propagation models can be used extensively in fields like Viral Marketing. Viral Marketing can be done by targeting a set of nodes in a network as done by Kleinberg et al in their paper on influence maximisation \cite{kempe2005influential,kempe2003maximizing}. Influential spreaders play a significant role in information propagation as shown by Kitsak et al in their work \cite{influence}. Meme virality can not only depend on network structure and nodes in a network, it seems to be intuitive that meme content also has a role to play in the meme becoming viral \cite{textviral,natures}. Though most studies consider nodes in their study of meme virality, we consider the property of edges in the spread of information. An edge connecting a “vulnerable” node to an influential node may have more impact as compared to a “vulnerable” node to another. Our study takes the diversity of edges into consideration and then probes into the meme pattern that can be formed.

\section{Generation of Networks for Meme Simulation : SCCP Networks}
It has been observed that most of the social networks are scale free and can be generated by the preferential attachment model. Further, these networks have communities because of the phenomenon of homophily that leads to the formation of dense clusters in the network. We also consider one more meso scale characteristic in the formation of network- core-periphery structure. It has been shown that the scale free networks possess an implicit core-periphery structure. Considering these 3 characteristics, we have tried to simulate real world networks via “SCCP” networks which show properties like \textbf{S}cale-free structure, presence of \textbf{C}ommunities and \textbf{C}ore-\textbf{P}eriphery structure. We introduce a modification to the algorithm \cite{cascades} employed by Wu et al. to generate these synthetic networks.

The algorithm is described below:\\

\textbf{Input:}
\begin{enumerate}
\item $k$= Number of communities in the network
\item $s$= Initial number of nodes in each community
\item $t_i$= Number of new nodes to be added in community $i$ where $ 0 \leq i \leq k-1 $
\item $t= max(t_i)$     $\forall i$
\item $f$= Fraction of edges each incoming node makes in its own community. $0.5 \leq f \leq 1$
\item $[r_1,r_2]$= Range for the number of edges an incoming node can make. $r_1$ and $r_2$ decides the density of a graph
\end{enumerate}
 
\vspace{1mm}
\textbf{Output:}
\begin{enumerate}
\item $G$= The SCCP network formed from input parameters specified above
\end{enumerate}

\begin{algorithm}[h!]
\caption{Algorithm for generating a SCCP network}
\label{CHalgorithm}
\begin{algorithmic}[1]
\State  We start the formation of the network G with k cliques each of size s.
\State  We perform t iterations. In each iteration, we add one node to every community except for the communities in which $t_i$ number of new nodes have already been added. Following two steps are performed in order to add a new node.
\begin{itemize}
\item Each newly arriving node chooses a random number $m$ in the range $[r_1,r_2]$. Then, it makes m edges with the already existing nodes.
\item The newly arrived node makes $fm$ edges with the nodes in its own community and $(1-f)m$ edges with the nodes in other communities. All the edges are created in a preferential attachment manner.
\end{itemize}
\State We detect core nodes in the generated graph by using k-shell decomposition algorithm \cite{kshell}. We declare the core nodes to form a separate community of their own. So the total number of communities in the network now becomes $k+1$.
\end{algorithmic}
\end{algorithm}

In our algorithm, we relax the condition of every node making equal number of edges on its arrival.
It is because, when a new person joins a social network, it is not compulsory for him/her to make a predefined number of friends. The number of friendships vary from person to person. Moreover, the above algorithm allows the formation of communities of different sizes.

\begin{figure}[h!]
\centering
\includegraphics[width=9 cm]{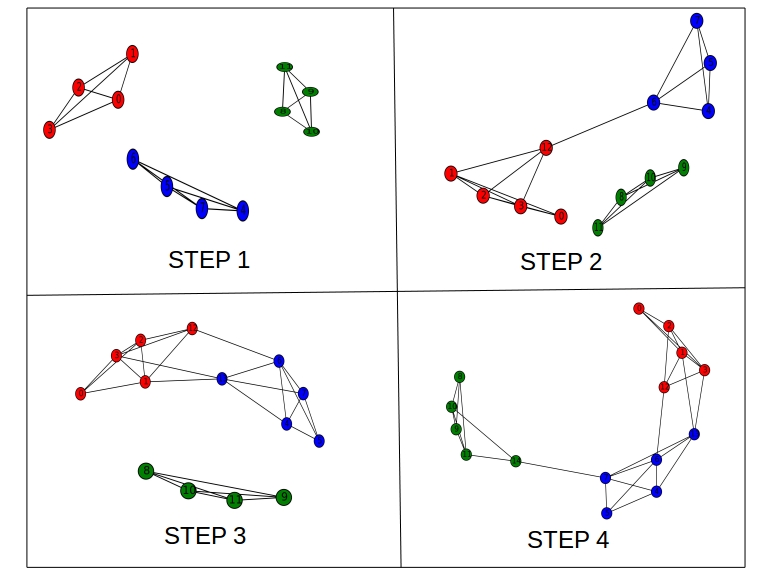}
\caption{Generation of a SCCP network}
\end{figure}

One iteration of this algorithm is illustrated in figure 1, where $k=3$, $s=4$, $t_i=1$ $\forall i$, $r1=2$, $r2=6$ , and $f=0.7$.
The network starts with 3 communities, each of which is a clique of 4 nodes. Communities 1, 2 and 3 are represented by red, blue and green colours respectively. Next, we add 3 nodes, one node to each community.
\begin{itemize}
\item First node is added in community 1. It chooses a random number 4. Then it makes 3 intra community edges and 1 inter community edge.
\item Second node is added in community 2. It chooses random number 5. Then it makes 3 intra community edges and 2 inter community edges.
\item The last node is added in community 3. It chooses random number 3. Then it makes 2 intra community edges and 1 inter community edge.  
\item Next, we detect core-periphery structure in the generated network by using k-shell decomposition
\end{itemize}

\section{Proposed Spreading Model}
Meme propagation on a real world network follows the pattern of a complex contagion. Unlike a simple contagion, the spreading pattern of a complex contagion depends on factors like homophily and social reinforcement\footnote{Homophily is the name given to the tendency of similar people becoming friends with each other. This leads to more number of ties between like minded people and hence leads to the formation of communities in the network. Social reinforcement is the phenomenon by which multiple exposures of an information to a person leads to him adopting it. Social reinforcement and homophily tend to block the information inside one community.}. A simple contagion is like an infectious disease which spreads with equal probability across all the edges, while a complex contagion spreads with different probabilities depending on the factors like social reinforcement and homophily \cite{natures}. In addition, user influence also plays a prominent role in meme propagation. We take into consideration all these factors in modeling the diffusion of a meme permeating through the ties in the network.\\

Our model is based on two key ideas :
\begin{enumerate}
\item \textit{Diversity in Tie Strength} : ``Birds of the same feather flock together''. We are more engaged and connected with the people in our own community as compared to people from other communities \cite{mcpherson2001birds}.
Hence, the probability associated with the edges connecting people of the same community should be higher than the edges connecting people of different communities. This observation gains motivation from the theory of weak ties \cite{granovetter1973strength}.
\item \textit{The social status of nodes} :
The social influence of a person in a network plays a big role in acceptance of information propagated by that person. A person's social status also decides if that person is vulnerable to adopting information. Simply stated, \textit{lower the status, higher the vulnerability} and vice versa. \textit{Higher the status, more the influence} and vice versa.
\end{enumerate}

Because of the presence of core-periphery structure in SCCP networks, there are two kinds of nodes in a SCCP network: core nodes and periphery nodes (periphery nodes are further divided into many communities). Initially, all the nodes are uninfected and a node turns infected as soon as it adopts a meme. We call an infected node $u$, the sender and an uninfected neighbour of $u$ say $v$, the receiver of an infection. The probability of infection transmission across an edge depends on the types of both nodes - the sender and the receiver. In our model, the probabilities of infection across edges are divided into five categories :
\begin{center}
$P_{cc}$, $P_{cp}$, $P_{pc}$, $P_{pp_0}$, $P_{pp_1}$ 
\end{center}
Here, `P' represents probability. The type of edge is represented by the subscript. The subscript's first alphabet denotes the type of sender node and second alphabet denotes the type of receiver node. `c' represents core, `p' represents periphery. 0 in the subscript denotes same community membership of sender and the receiver node, while 1 represents sender and receiver belonging to different communities. 
We worked towards predicting the most plausible order for these edge probabilities, which is initially proposed to be as :
\textit{$P_{cc} > P_{cp} > P_{pp_0} > P_{pp_1} > P_{pc} $}.\\
Our model can be considered as an extension of the simple cascade model, with a slight change in the definition in every iteration, each infected node tries infecting its uninfected neighbours in accordance with the above probability hierarchy.\\

\section{Datasets}
We have used multiple SCCP networks, random graphs \cite{erdHos1961strength} and real world networks \cite{snapnets} in our study. We have considered the two most widely used online social networks- Facebook and Twitter having approximately 1371 and 271 million users. For comparing our complete framework, we use the Higgs boson meme propagation information on Twitter (dataset1). The dataset 1 gives a complete picture of a meme spreading on an online social network along with the information ``who infected whom” at every step.The datasets' specifications are given below:
\begin{enumerate}
\item \textbf{Dataset 1:}
\textbf{Dataset 1(a):} This dataset is an induced directed unweighted subgraph on Twitter users who were involved in any of the activities(reply, retweet, or mention) regarding the Higgs boson meme\footnote{Higgs boson is one of the most elementary elusive particle in modern physics. A meme in Twitter is considered to be a Higgs Boson meme if it contains at least one of these keywords or tags: lhc, cern, boson, higgs} \cite{de2013anatomy}. It is an undirected unweighted graph containing 456631 nodes and 14855875 edges.\\
\textbf{Dataset 1(b):} This is a directed weighted graph between the Twitter users who were involved in retweeting \cite{yang2010understanding} of the Higgs Boson meme. There is an edge from B to A if A retweets B. This graph contains 425008 nodes and 733647 edges.
In datasets 1(a) and 1(b), the tweets posted in Twitter about this discovery between 1st and 7th July 2012 are considered.
\item \textbf{Dataset 2}: This dataset is an undirected unweighted induced subgraph on Facebook with 4039 nodes and 88234 edges \cite{leskovec2012learning}.  
\item \textbf{Dataset 3}: This dataset is an induced undirected unweighted subgraph on Twitter with 81306 nodes and 1768149 edges  \cite{leskovec2012learning}.
\item \textbf{Dataset 4}: These datasets have been derived from the algorithm proposed in the previous section.\\
\textbf{Dataset 4(a):} This is a SCCP network on $65800$ nodes, $591750$ edges and $11$ communities.\\
\textbf{Dataset 4(b):} This is a SCCP network on $4000$ nodes, $170314 $ edges and $11$ communities.
\item \textbf{Dataset 5}: This is an Erdos-Renyi graph on $4000$ nodes and $34650$ edges.
\end{enumerate}

We detect communities in datasets 1(a), 2 and 3 using fast greedy modularity optimization algorithm. This algorithm is given by Newman et. al. \cite{clauset2004finding} and is used to detect community structure for very large graphs. We also find out the core-periphery structure for all the above listed datasets using k-shell decomposition algorithm. We assign a coreness value to each node equal to the shell value assigned to it by the algorithm. Then, we pick top 10\% of the nodes having highest coreness values and call them the core nodes. The remainder of the nodes are termed periphery nodes.\\

\section{Experiments and Results}
\subsection{Spreading Model Validation}
Our model was validated using datasets 1(a) and 1(b), where 1(a) gives us the information about the structure of a social network and 1(b) is the cascading pattern of a meme over 1(a). Let the dataset 1(a) be represented by $G(V,E)$.
Based on the structure of $G$, we partition its nodes in two subsets $C$ and $P$. $C$ is the set of core nodes and $P$ is the set of periphery nodes such that $C \cup P = V $ and $ C \cap P = \emptyset$. We also associate a variable $\delta_{ij}$ with each edge $E_{ij}$. $\delta_{ij} = 1$ if nodes $i$ and $j$ belong to the same community, else 0.   
We divide the edges in the retweet network (dataset 1(b)) in four categories based on the types of users an edge is connecting. These categories are as follows:-
\begin{enumerate}
\item $E_{cc}$ = $\{E_{ij} \in E :  (i \in C) \wedge ( j \in C)\}$
\item $E_{cp}$ = $\{E_{ij} \in E :  (i \in C) \wedge ( j \in P)\}$
\item $E_{pc}$ = $\{E_{ij} \in E  : (i \in P) \wedge ( j \in C)\}$
\item $E_{pp}$ = $\{E_{ij} \in E :  (i \in P) \wedge ( j \in P)\}$
\begin{itemize}
\item $E_{{pp}_0}$ = $\{E_{ij} \in E : (i \in P) \wedge ( j \in P) \wedge \delta_{ij} = 1\}$
\item $E_{{pp}_1}$ = $ \{E_{ij} \in E : (i \in P) \wedge ( j \in P) \wedge \delta_{ij} = 0\}$
\end{itemize}
\end{enumerate}
The types of nodes for 1(b) are extracted from its main graph 1(a).\\

In retweet networks, the weight of an edge from A to B specifies the amount of information flowing from A to B (number of times B retweeted a message from A). Therefore, more the weight, higher the probability of information transmission across that edge.  
We calculate the following weights from the above graphs:-\\
Let $W(E_{ij})$ be the weight of an edge from node $i$ to node $j$ and $N_{xy}$ represent the type of edges $E_{xy}$ where $x$ and $y$ are the types of nodes hence having the possible values $p$ and $c$.
Then, we calculate $W_{xy}$ ,the sum of weights of all the edges from a node of type $x$ to a node of type $y$. 
 \begin{enumerate}
\item $W_{cc}$ = $ \sum (W(E_{ij})  )/  N_{cc}  $ such that  $ E_{ij} \in E_{cc}$
\item $W_{cp}$ = $ \sum  (W(E_{ij}) )/  N_{cp}$ such that  $ E_{ij} \in E_{cp}$
\item $W_{pc}$ = $ \sum  (W(E_{ij}) )/  N_{pc}$ such that  $ E_{ij} \in E_{pc}$
\item $W_{pp}$ = $ \sum  (W(E_{ij}) )/  N_{pp} \mid$ such that  $ E_{ij} \in E_{pp}$
 \begin{itemize}
\item $W_{{pp}_0}$ = $ \sum  (W(E_{ij} ) )/  N_{pp_0}$ such that   $ E_{ij} \in E_{pp_0}$
\item $W_{{pp}_1}$ = $ \sum  (W(E_{ij} ))/  N_{pp_1}$ such that   $ E_{ij} \in E_{pp_1}$
\end{itemize}
\end{enumerate}
The weights obtained show that the observed order is the same as we have proposed earlier thereby validating the ordering we proposed i.e.
$W_{cc} > W_{cp} > W_{{pp}_0} > W_{{pp}_1} > W_{pc} $.

\subsection{Simulation Results}
We introspect on the extent as well as rate of infection of the network, while propagating a meme on it. We simulate EBH as well as uniform spreading model on a number of datasets and report the results. For the simulation of our proposed model, we use the following probabilities:
 $E_{pc}:0.00001$,
 $E_{{pp}_0}:0.0003$,
 $E_{{pp}_0}:0.0001$,
 $E_{cc}:0.006$, and
 $E_{cp}:0.004$.
For the simulation of uniform spreading model, every edge is considered to have an equal probability of infection i.e. $E_{ij}=0.0002$, where i and j are the endpoints of an edge. We have chosen these probabilities such that we can visualise the spreading pattern of a meme to the best possible extent. For all the figures in this section, X axis represents the number of iterations and Y axis represents the cumulative number of nodes infected up to that iteration. The results of this paper are structured in three parts :

\begin{figure}
\includegraphics[width= 0.3\textwidth]{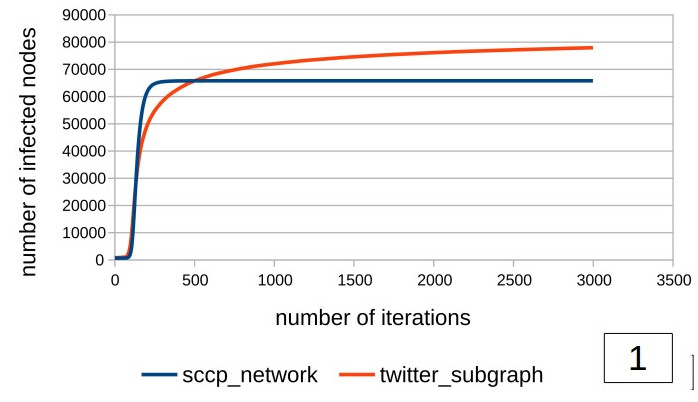}
 \includegraphics[width= 0.3\textwidth]{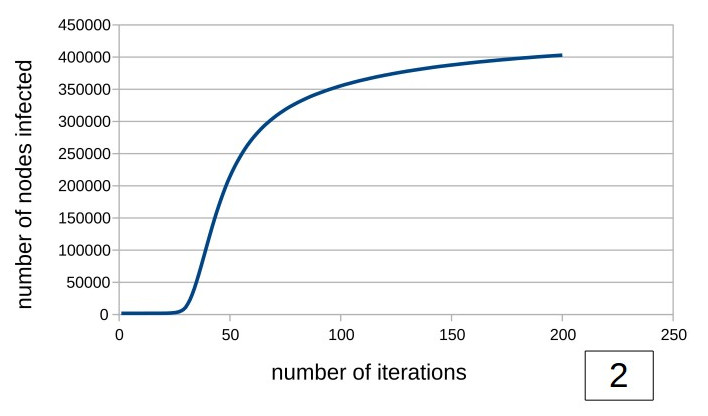}
\includegraphics[width= 0.3\textwidth]{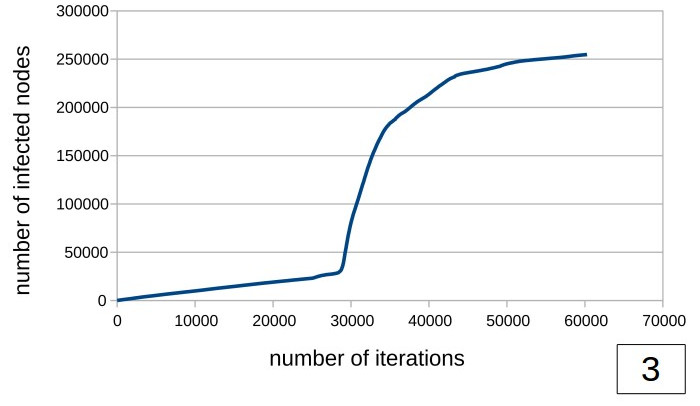}
\includegraphics[width= 0.3\textwidth]{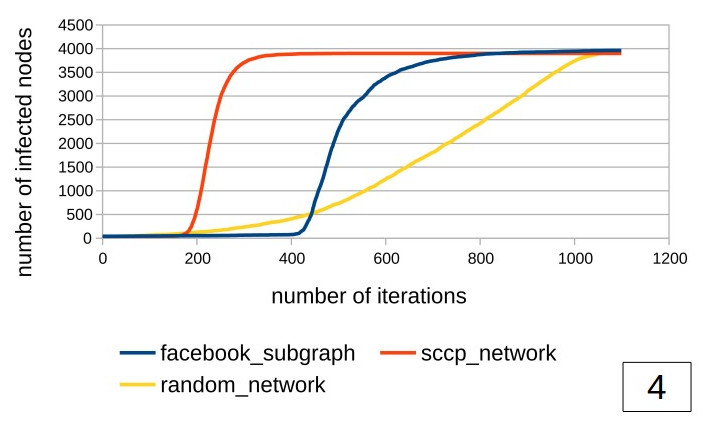} \hspace{4mm}
\includegraphics[width= 0.3\textwidth]{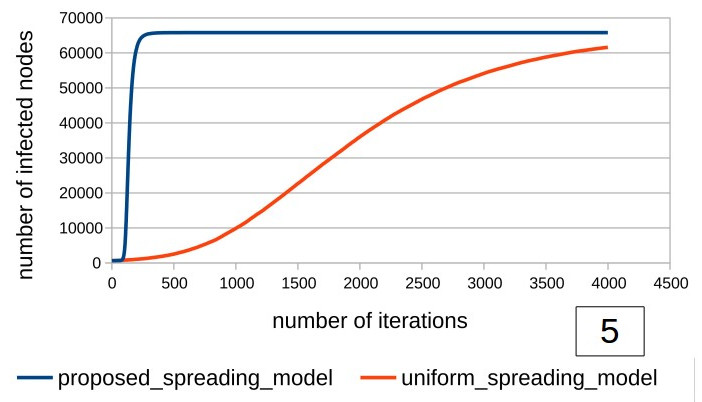} \hspace{4mm}
\includegraphics[width= 0.3\textwidth]{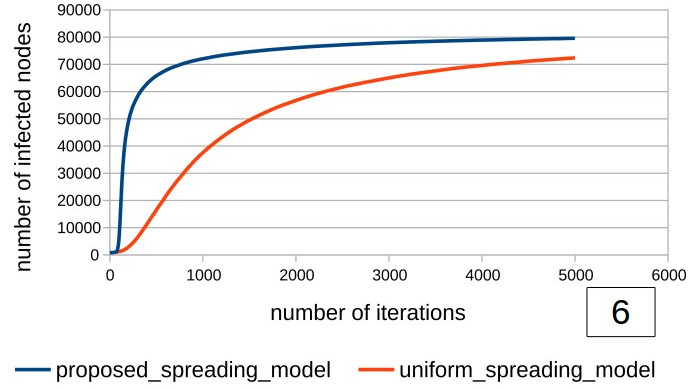}
 
\caption{Spreading patterns on different kinds of networks and its comparison to real world data - \textit{1:} Spreading patterns on datasets 3 and 4(a) \textit{2:} Spreading pattern on dataset 1(a) \textit{3:} Actual spreading pattern of Higgs boson meme(dataset 1(a) and 1(b)) \textit{4:} Comparison between the proposed spreading model on datasets 2, 4(a), and 5 \textit{5:} Spreading patterns for dataset 4(a) \textit{6:} Proposed and uniform spreading models on dataset 3}
\end{figure}

\begin{figure}
\includegraphics[width= 0.4\textwidth]{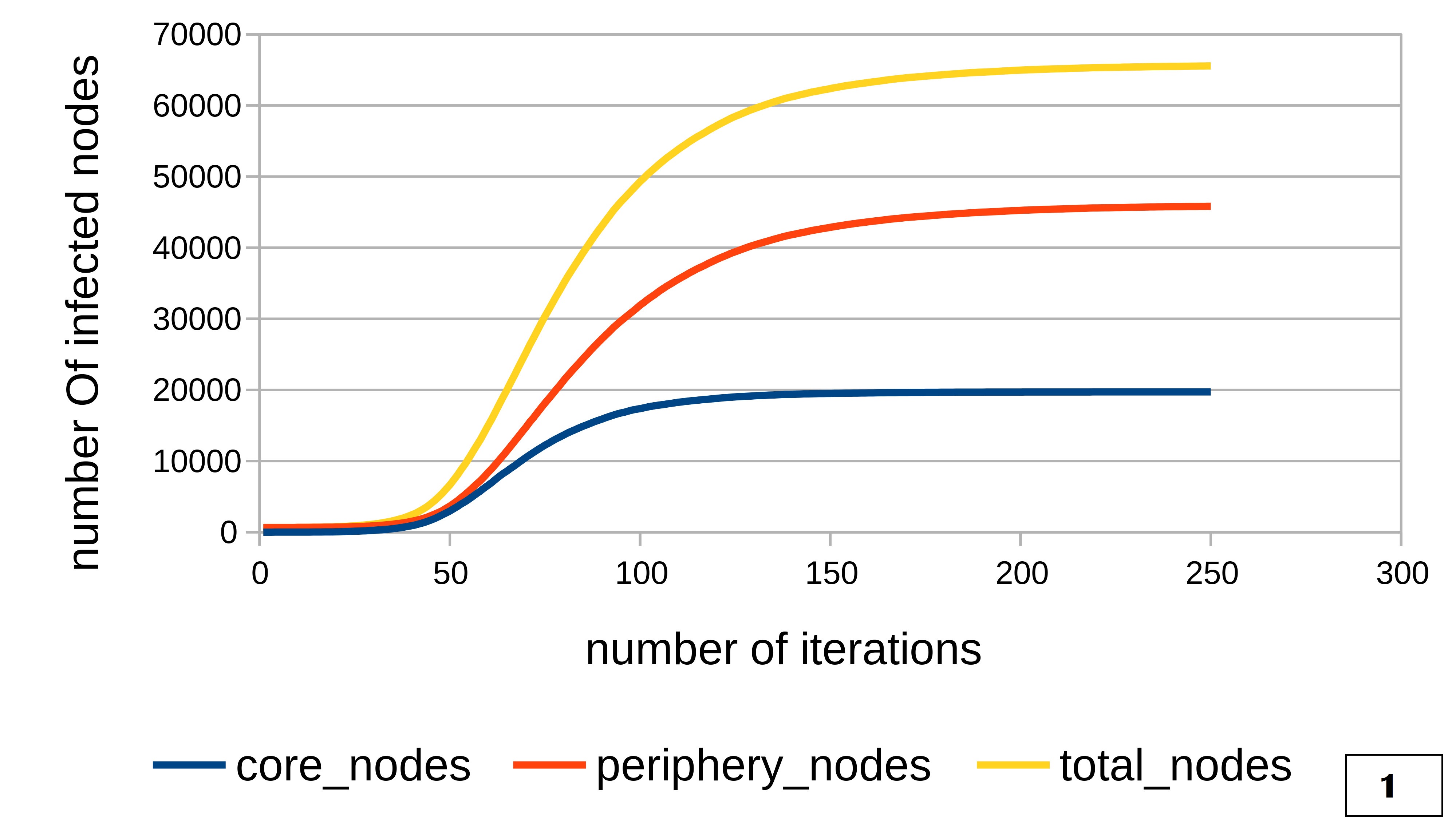}
 \includegraphics[width= 0.4\textwidth]{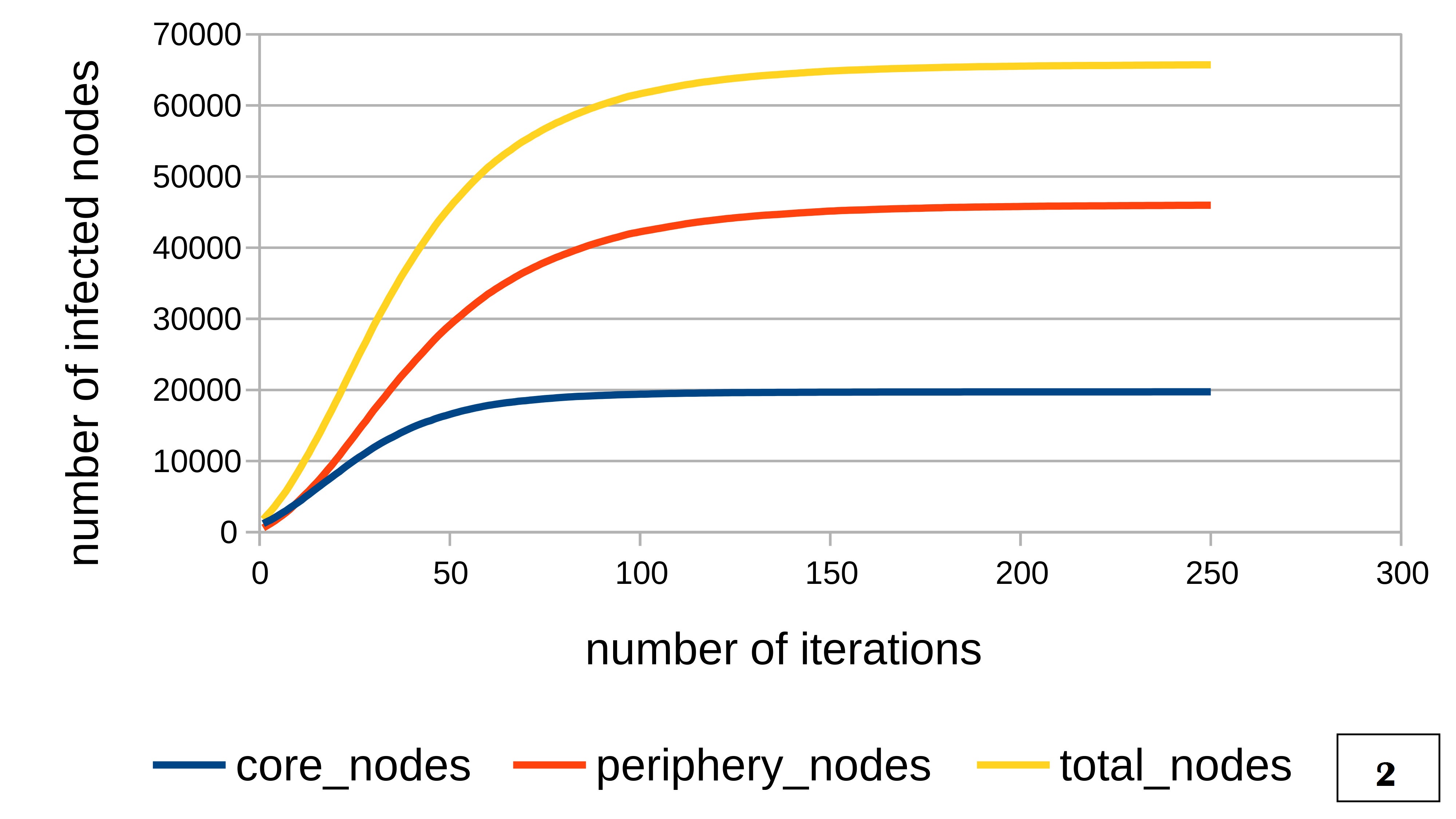}
 \includegraphics[width= 0.4\textwidth]{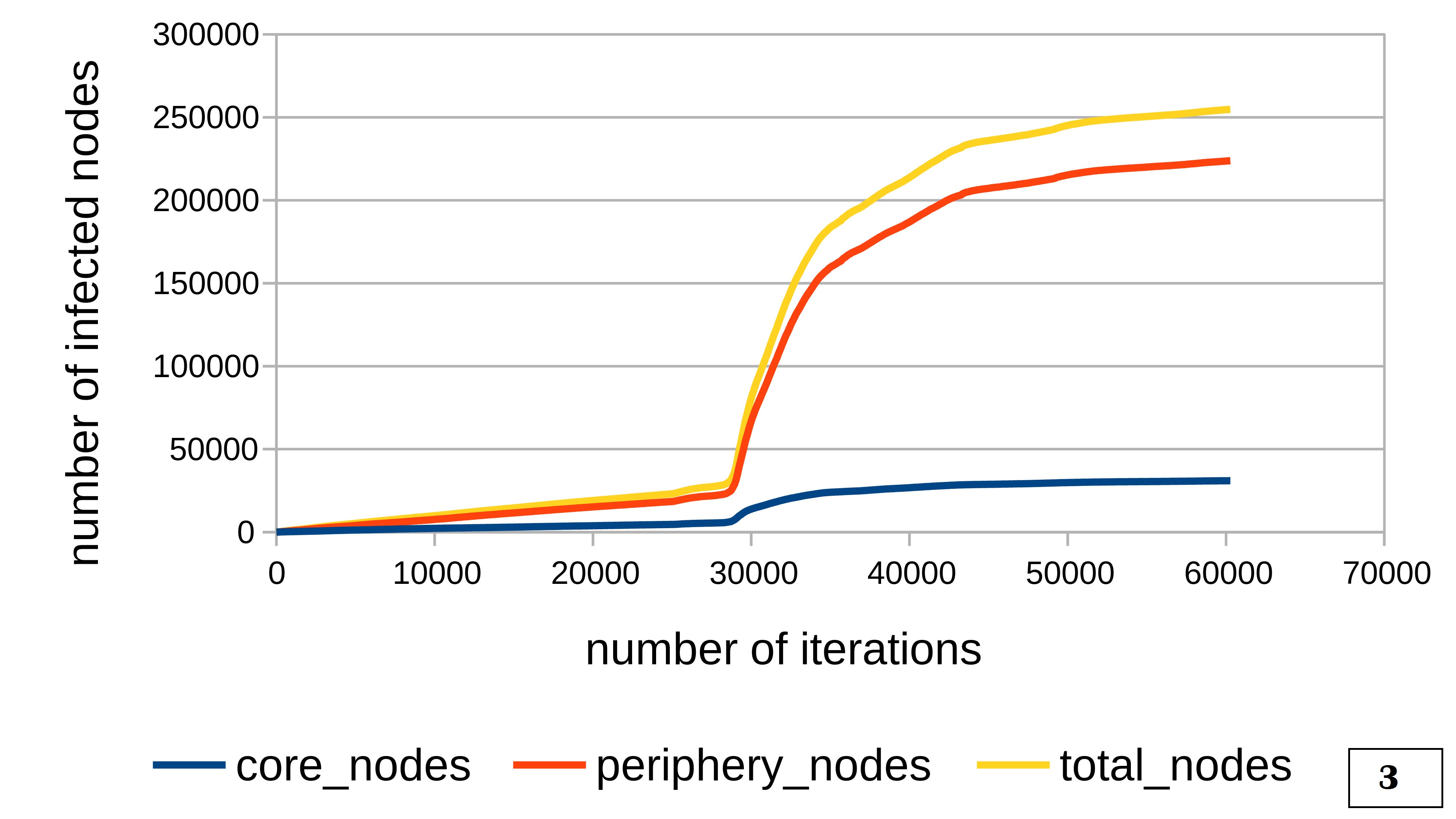}
 \includegraphics[width= 0.5\textwidth]{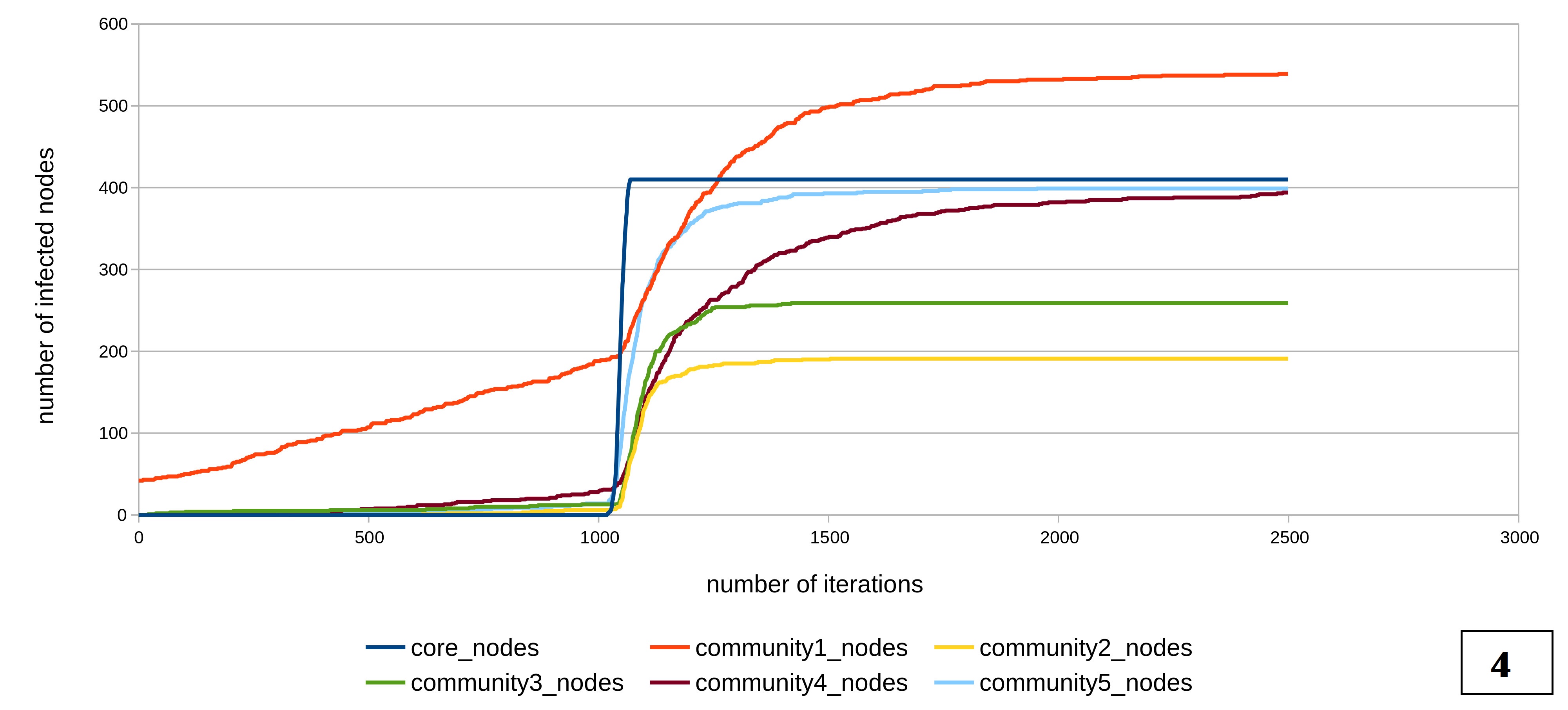}
 \includegraphics[width= 0.5\textwidth]{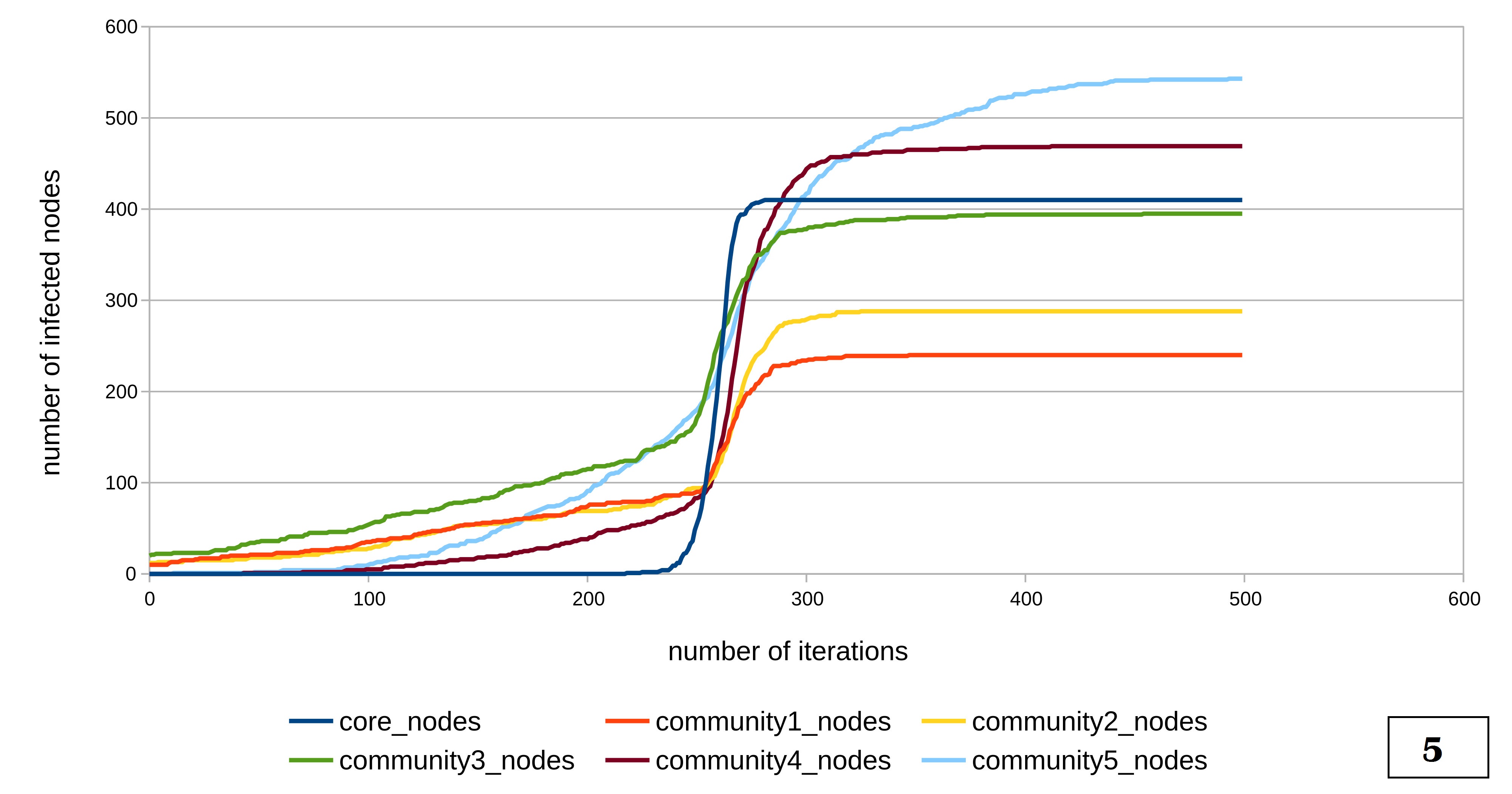}
  \includegraphics[width= 0.5\textwidth]{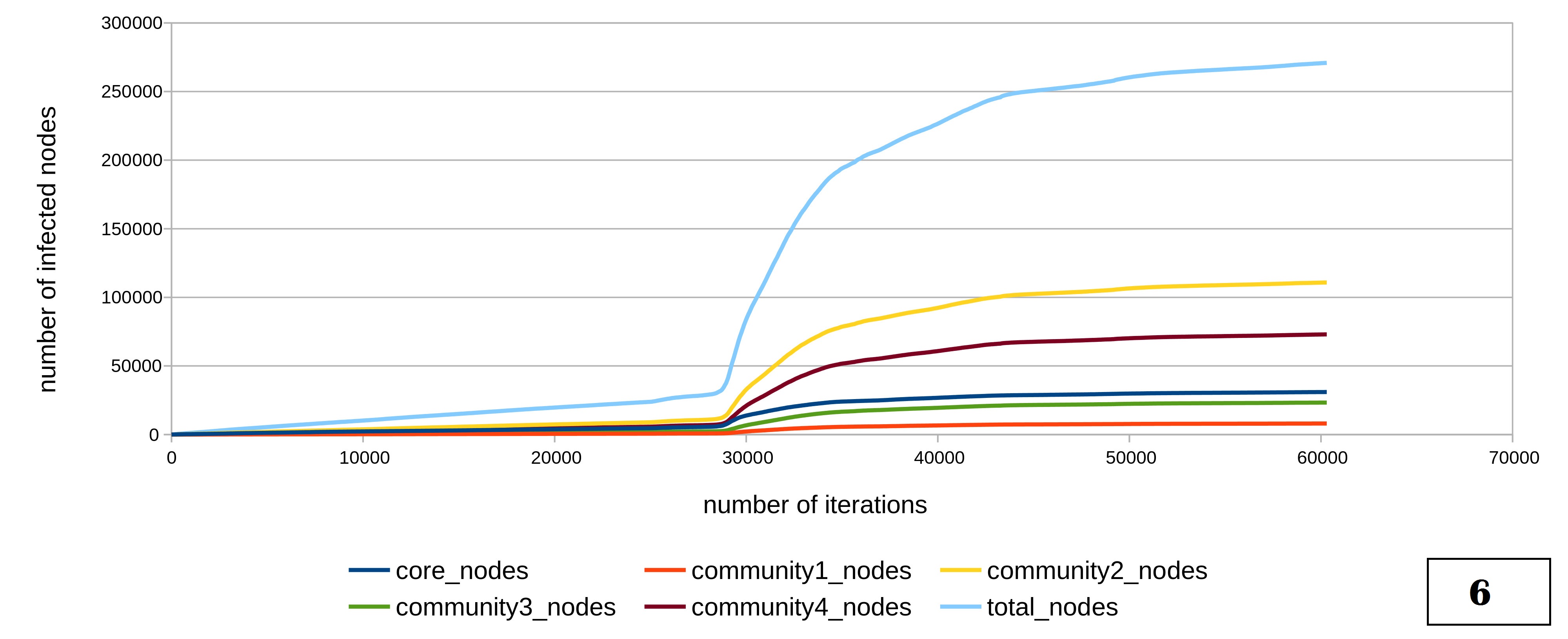}
  \caption{Spreading patterns starting from different types of seed nodes and its comparison to real world data -\textit{1}: Proposed spreading model on dataset 4(a) where spreading starts from periphery nodes  \textit{2}: Proposed spreading model on dataset 4(a) where spreading starts from core nodes  \textit{3}: Actual spreading pattern for the Higgs boson meme(from dataset 1(a) and 1(b)) \textit{4}: Spreading patterns starting from single community \textit{5}:Spreading patterns starting from multiple communities  \textit{6}: Actual spreading pattern for the Higgs boson meme(from dataset 1(a) and 1(b))}
 \end{figure}
 
\subsubsection{Meme spreading patterns on different networks using the EBH and uniform spreading models}
Figure 1.2 shows the actual spreading pattern of the Higgs boson meme which indicates that in the real world, a meme does not have a constant growth rate.  The rate remains constant upto some point, after which the popularity of a meme shoots up steeply and then slowly fades, giving rise to a sigmoid curve which is characterised by the equation :
$F(x)= 1/(1+e^{-kx})$
Figure 1.1 shows the simulation of our proposed spreading model on the SCCP network and two real world networks of Facebook and Twitter. It can be seen that in both these cases, the curve for the spreading pattern is seen to be sigmoidal just like figure 1.2. 
Figure 1.4 shows the difference in the spreading patterns when the simulation is done through an uniform spreading model and our model respectively. It can be seen that the simulation through an uniform spreading model is also a sigmoid function but has a lesser value of parameter x.
Figure 1.3 shows the simulation of the proposed spreading model on 3 different kinds of networks. Despite simulating the EBH spreading model on all the three graphs, the value of x is observed to be lower only in the case of random networks\footnote{In the case of random network, even though the declared 10\% core nodes have a high probability of infecting their neighbours, the connections between the core nodes are not dense enough to result in an overshoot in the number of infected nodes. So, absence of a distinct core-periphery structure in such networks make them invalid for our framework} Thus we can say that the sharp S shaped infection pattern is observed only for the SCCP kind of networks. \textit{These graphs show that the presence of both- a SCCP kind of network as well as EBH spreading model are required to mimic a real world meme propagation. } Some similar results have been described in the attached appendix B.

\subsubsection{Explanation of the plateau structure observed in the meme pattern}
Figure 2.3 shows the pattern of infection of core nodes and periphery nodes for the actual Higgs boson meme. As in the previous case, all iterations are considered to be of equal length (10 timestamps). We observe the cumulative number of core nodes and periphery nodes infected in every iteration. When we started infection from periphery nodes(figure 2.1), the plateau structure of the curve continues till a core node is infected and then the infection shoots up suddenly. Figure 2.2 shows the plot when the infection is started only from the core nodes. We can see that in this case, infection shoots up immediately without the plateau structure. \textit{This solidifies the observation that the number of periphery nodes infected increases sharply as soon as a sufficient fraction of the core nodes gets infected.}

\subsubsection{Effect of communities and core nodes on meme virality}
In figure 2.4, we start the infection from a single community and show that the infection spreads in multiple communities only when the meme infects the core sufficiently and gets viral. Figure 2.5 shows the spreading pattern when the infection starts from multiple communities. But the meme becomes viral only after the infection of core nodes. So, \textit{whether the infection starts from single community or multiple communities, the infection of core nodes is sufficient to predict its virality}. Figure 2.6 shows the actual spreading pattern of Higgs boson meme. 

\section{Conclusion and Future Work}

A lot of researchers are working towards proposing the models that can predict the pattern of meme spread in a real world network today. A number of models have been proposed for this ranging from simple epidemiological models to the standard models like Linear Threshold and Independent Cascade. Most of these models do not give an approach to identify the parameters required to simulate them. Moreover, they are proposed for all kind of networks though they can be improved upon and specialised for a particular kind of network. Hence, improving these models to better simulate a meme propagation is possible. It is shown that, together, SCCP and EBH models effectively simulate real world meme propagation. The sigmoid curve with a sharp slope is shown to be the characteristic pattern of an internet meme. Furthermore, the importance of core nodes in marking the virality of a meme is emphasised. It is also shown that infecting multiple communities also require the infection of core nodes. The study is validated with the Higgs boson meme spreading on Twitter in addition to various other real world networks. This study opens a new direction of considering edge diversity in meme propagation models. \\
One can extend our problem to predict the exact values of the probabilities influencing the meme propagation. This can greatly help in prediction of a future cascade pattern. If such cascades could be predetermined then we could exert a control on our otherwise ever changing social networks. Not only could preventive checkpoints be placed in the network but also useful information could be accelerated through the network by using the predicted meme trajectory.

\bibliographystyle{IEEEtran}
\bibliography{mybib}

\begin{thebibliography}{10}
\providecommand{\url}[1]{#1}
\csname url@samestyle\endcsname
\providecommand{\newblock}{\relax}
\providecommand{\bibinfo}[2]{#2}
\providecommand{\BIBentrySTDinterwordspacing}{\spaceskip=0pt\relax}
\providecommand{\BIBentryALTinterwordstretchfactor}{4}
\providecommand{\BIBentryALTinterwordspacing}{\spaceskip=\fontdimen2\font plus
\BIBentryALTinterwordstretchfactor\fontdimen3\font minus
  \fontdimen4\font\relax}
\providecommand{\BIBforeignlanguage}[2]{{%
\expandafter\ifx\csname l@#1\endcsname\relax
\typeout{** WARNING: IEEEtran.bst: No hyphenation pattern has been}%
\typeout{** loaded for the language `#1'. Using the pattern for}%
\typeout{** the default language instead.}%
\else
\language=\csname l@#1\endcsname
\fi
#2}}
\providecommand{\BIBdecl}{\relax}
\BIBdecl

\bibitem{clissold2004candid}
B.~D. Clissold, ``Candid camera and the origins of reality tv,''
  \emph{Understanding reality television}, pp. 33--53, 2004.

\bibitem{christakis2007spread}
N.~A. Christakis and J.~H. Fowler, ``The spread of obesity in a large social
  network over 32 years,'' \emph{New England journal of medicine}, vol. 357,
  no.~4, pp. 370--379, 2007.

\bibitem{statisticsstatistic}
S.~N. Statistics, ``Statistic brain (july 9, 2014).''

\bibitem{daley1964epidemics}
D.~J. Daley and D.~G. Kendall, ``Epidemics and rumours,'' 1964.

\bibitem{hethcote1989three}
H.~W. Hethcote, ``Three basic epidemiological models,'' in \emph{Applied
  mathematical ecology}.\hskip 1em plus 0.5em minus 0.4em\relax Springer, 1989,
  pp. 119--144.

\bibitem{goldenberg2001talk}
J.~Goldenberg, B.~Libai, and E.~Muller, ``Talk of the network: A complex
  systems look at the underlying process of word-of-mouth,'' \emph{Marketing
  letters}, vol.~12, no.~3, pp. 211--223, 2001.

\bibitem{granovetter1978threshold}
M.~Granovetter, ``Threshold models of collective behavior,'' \emph{American
  journal of sociology}, pp. 1420--1443, 1978.

\bibitem{bakshy2012role}
E.~Bakshy, I.~Rosenn, C.~Marlow, and L.~Adamic, ``The role of social networks
  in information diffusion,'' in \emph{Proceedings of the 21st international
  conference on World Wide Web}.\hskip 1em plus 0.5em minus 0.4em\relax ACM,
  2012, pp. 519--528.

\bibitem{digg}
K.~Lerman and R.~Ghosh, ``Information contagion: An empirical study of the
  spread of news on digg and twitter social networks.'' \emph{ICWSM}, vol.~10,
  pp. 90--97, 2010.

\bibitem{kleinberg2007cascading}
J.~Kleinberg, ``Cascading behavior in networks: Algorithmic and economic
  issues,'' \emph{Algorithmic game theory}, vol.~24, pp. 613--632, 2007.

\bibitem{saito2008prediction}
K.~Saito, R.~Nakano, and M.~Kimura, ``Prediction of information diffusion
  probabilities for independent cascade model,'' in \emph{Knowledge-based
  intelligent information and engineering systems}.\hskip 1em plus 0.5em minus
  0.4em\relax Springer, 2008, pp. 67--75.

\bibitem{pastor2001epidemic}
R.~Pastor-Satorras and A.~Vespignani, ``Epidemic spreading in scale-free
  networks,'' \emph{Physical review letters}, vol.~86, no.~14, p. 3200, 2001.

\bibitem{zhang2011research}
Z.~Y.-C. L.~Y. Zhang and H.-F. C. H.~X. Fei, ``The research of information
  dissemination model on online social network [j],'' \emph{Acta Physica
  Sinica}, vol.~5, p. 010, 2011.

\bibitem{budak2011limiting}
C.~Budak, D.~Agrawal, and A.~El~Abbadi, ``Limiting the spread of misinformation
  in social networks,'' in \emph{Proceedings of the 20th international
  conference on World wide web}.\hskip 1em plus 0.5em minus 0.4em\relax ACM,
  2011, pp. 665--674.

\bibitem{scanfeld2010dissemination}
D.~Scanfeld, V.~Scanfeld, and E.~L. Larson, ``Dissemination of health
  information through social networks: Twitter and antibiotics,''
  \emph{American journal of infection control}, vol.~38, no.~3, pp. 182--188,
  2010.

\bibitem{textviral}
M.~Guerini, C.~Strapparava, and G.~{\"O}zbal, ``Exploring text virality in
  social networks.'' in \emph{ICWSM}, 2011.

\bibitem{natures}
L.~Weng, F.~Menczer, and Y.-Y. Ahn, ``Virality prediction and community
  structure in social networks,'' \emph{Scientific reports}, vol.~3, 2013.

\bibitem{kempe2005influential}
D.~Kempe, J.~Kleinberg, and {\'E}.~Tardos, ``Influential nodes in a diffusion
  model for social networks,'' in \emph{Automata, languages and
  programming}.\hskip 1em plus 0.5em minus 0.4em\relax Springer, 2005, pp.
  1127--1138.

\bibitem{kempe2003maximizing}
------, ``Maximizing the spread of influence through a social network,'' in
  \emph{Proceedings of the ninth ACM SIGKDD international conference on
  Knowledge discovery and data mining}.\hskip 1em plus 0.5em minus 0.4em\relax
  ACM, 2003, pp. 137--146.

\bibitem{influence}
M.~Kitsak, L.~K. Gallos, S.~Havlin, F.~Liljeros, L.~Muchnik, H.~E. Stanley, and
  H.~A. Makse, ``Identification of influential spreaders in complex networks,''
  \emph{Nature Physics}, vol.~6, no.~11, pp. 888--893, 2010.

\bibitem{cascades}
J.-j. Wu, Z.-y. Gao, and H.-j. Sun, ``Cascade and breakdown in scale-free
  networks with community structure,'' \emph{Physical Review E}, vol.~74,
  no.~6, p. 066111, 2006.

\bibitem{della}
F.~Della~Rossa, F.~Dercole, and C.~Piccardi, ``Profiling core-periphery network
  structure by random walkers,'' \emph{Scientific reports}, vol.~3, 2013.

\bibitem{mcpherson2001birds}
M.~McPherson, L.~Smith-Lovin, and J.~M. Cook, ``Birds of a feather: Homophily
  in social networks,'' \emph{Annual review of sociology}, pp. 415--444, 2001.

\bibitem{granovetter1973strength}
M.~S. Granovetter, ``The strength of weak ties,'' \emph{American journal of
  sociology}, pp. 1360--1380, 1973.

\bibitem{clauset2004finding}
A.~Clauset, M.~E. Newman, and C.~Moore, ``Finding community structure in very
  large networks,'' \emph{Physical review E}, vol.~70, no.~6, p. 066111, 2004.

\bibitem{de2013anatomy}
M.~De~Domenico, A.~Lima, P.~Mougel, and M.~Musolesi, ``The anatomy of a
  scientific rumor,'' \emph{Scientific reports}, vol.~3, 2013.

\bibitem{yang2010understanding}
Z.~Yang, J.~Guo, K.~Cai, J.~Tang, J.~Li, L.~Zhang, and Z.~Su, ``Understanding
  retweeting behaviors in social networks,'' in \emph{Proceedings of the 19th
  ACM international conference on Information and knowledge management}.\hskip
  1em plus 0.5em minus 0.4em\relax ACM, 2010, pp. 1633--1636.

\bibitem{leskovec2012learning}
J.~Leskovec and J.~J. Mcauley, ``Learning to discover social circles in ego
  networks,'' in \emph{Advances in neural information processing systems},
  2012, pp. 539--547.

\bibitem{erdHos1961strength}
P.~Erd{\H{o}}s and A.~R{\'e}nyi, ``On the strength of connectedness of a random
  graph,'' \emph{Acta Mathematica Hungarica}, vol.~12, no.~1, pp. 261--267,
  1961.

\bibitem{kshell}
V.~Batagelj and M.~Zaversnik, ``An o (m) algorithm for cores decomposition of
  networks,'' \emph{arXiv preprint cs/0310049}, 2003.

\bibitem{snapnets}
J.~Leskovec and A.~Krevl, ``{SNAP Datasets}: {Stanford} large network dataset
  collection,'' Jun. 2014.

\end{thebibliography}


\begin{thebibliography}{10}
\providecommand{\url}[1]{#1}
\csname url@samestyle\endcsname
\providecommand{\newblock}{\relax}
\providecommand{\bibinfo}[2]{#2}
\providecommand{\BIBentrySTDinterwordspacing}{\spaceskip=0pt\relax}
\providecommand{\BIBentryALTinterwordstretchfactor}{4}
\providecommand{\BIBentryALTinterwordspacing}{\spaceskip=\fontdimen2\font plus
\BIBentryALTinterwordstretchfactor\fontdimen3\font minus
  \fontdimen4\font\relax}
\providecommand{\BIBforeignlanguage}[2]{{%
\expandafter\ifx\csname l@#1\endcsname\relax
\typeout{** WARNING: IEEEtran.bst: No hyphenation pattern has been}%
\typeout{** loaded for the language `#1'. Using the pattern for}%
\typeout{** the default language instead.}%
\else
\language=\csname l@#1\endcsname
\fi
#2}}
\providecommand{\BIBdecl}{\relax}
\BIBdecl

\bibitem{daley1964epidemics}
D.~J. Daley and D.~G. Kendall, ``Epidemics and rumours,'' 1964.

\bibitem{mast}
W.~GOFFMAN, ``Mathematical approach to the spread of scientific ideas- the
  history of mast cell research,'' \emph{Nature}, vol. 212, 1966.

\bibitem{tabah1999literature}
A.~N. Tabah, ``Literature dynamics: Studies on growth, diffusion, and
  epidemics.'' \emph{Annual review of information science and technology
  (ARIST)}, vol.~34, pp. 249--86, 1999.

\bibitem{cheng2014can}
J.~Cheng, L.~Adamic, P.~A. Dow, J.~M. Kleinberg, and J.~Leskovec, ``Can
  cascades be predicted?'' in \emph{Proceedings of the 23rd international
  conference on World wide web}.\hskip 1em plus 0.5em minus 0.4em\relax ACM,
  2014, pp. 925--936.

\bibitem{leskovec2007patterns}
J.~Leskovec, M.~McGlohon, C.~Faloutsos, N.~S. Glance, and M.~Hurst, ``Patterns
  of cascading behavior in large blog graphs.'' in \emph{SDM}, vol.~7.\hskip
  1em plus 0.5em minus 0.4em\relax SIAM, 2007, pp. 551--556.

\bibitem{ba}
A.-L. Barab{\'a}si and R.~Albert, ``Emergence of scaling in random networks,''
  \emph{science}, vol. 286, no. 5439, pp. 509--512, 1999.

\bibitem{isaak2004globalization}
R.~A. Isaak, \emph{The globalization gap: How the rich get richer and the poor
  get left further behind}.\hskip 1em plus 0.5em minus 0.4em\relax Pearson
  Education, 2004.

\bibitem{fortunato2010community}
S.~Fortunato, ``Community detection in graphs,'' \emph{Physics Reports}, vol.
  486, no.~3, pp. 75--174, 2010.

\bibitem{girvan2002community}
M.~Girvan and M.~E. Newman, ``Community structure in social and biological
  networks,'' \emph{Proceedings of the National Academy of Sciences}, vol.~99,
  no.~12, pp. 7821--7826, 2002.

\bibitem{borgatti2000models}
S.~P. Borgatti and M.~G. Everett, ``Models of core/periphery structures,''
  \emph{Social networks}, vol.~21, no.~4, pp. 375--395, 2000.

\bibitem{della}
F.~Della~Rossa, F.~Dercole, and C.~Piccardi, ``Profiling core-periphery network
  structure by random walkers,'' \emph{Scientific reports}, vol.~3, 2013.

\bibitem{kumar2010structure}
R.~Kumar, J.~Novak, and A.~Tomkins, ``Structure and evolution of online social
  networks,'' in \emph{Link mining: models, algorithms, and
  applications}.\hskip 1em plus 0.5em minus 0.4em\relax Springer, 2010, pp.
  337--357.

\bibitem{borzsei2013makes}
L.~K. B{\"o}rzsei, ``Makes a meme instead: A concise history of internet
  memes,'' \emph{New Media Studies Magazine}, no.~7, 2013.

\bibitem{cascades}
J.-j. Wu, Z.-y. Gao, and H.-j. Sun, ``Cascade and breakdown in scale-free
  networks with community structure,'' \emph{Physical Review E}, vol.~74,
  no.~6, p. 066111, 2006.

\bibitem{kshell}
V.~Batagelj and M.~Zaversnik, ``An o (m) algorithm for cores decomposition of
  networks,'' \emph{arXiv preprint cs/0310049}, 2003.

\bibitem{erdHos1961strength}
P.~Erd{\H{o}}s and A.~R{\'e}nyi, ``On the strength of connectedness of a random
  graph,'' \emph{Acta Mathematica Hungarica}, vol.~12, no.~1, pp. 261--267,
  1961.

\bibitem{snapnets}
J.~Leskovec and A.~Krevl, ``{SNAP Datasets}: {Stanford} large network dataset
  collection,'' Jun. 2014.

\bibitem{de2013anatomy}
M.~De~Domenico, A.~Lima, P.~Mougel, and M.~Musolesi, ``The anatomy of a
  scientific rumor,'' \emph{Scientific reports}, vol.~3, 2013.

\bibitem{yang2010understanding}
Z.~Yang, J.~Guo, K.~Cai, J.~Tang, J.~Li, L.~Zhang, and Z.~Su, ``Understanding
  retweeting behaviors in social networks,'' in \emph{Proceedings of the 19th
  ACM international conference on Information and knowledge management}.\hskip
  1em plus 0.5em minus 0.4em\relax ACM, 2010, pp. 1633--1636.

\bibitem{leskovec2012learning}
J.~Leskovec and J.~J. Mcauley, ``Learning to discover social circles in ego
  networks,'' in \emph{Advances in neural information processing systems},
  2012, pp. 539--547.

\bibitem{clauset2004finding}
A.~Clauset, M.~E. Newman, and C.~Moore, ``Finding community structure in very
  large networks,'' \emph{Physical review E}, vol.~70, no.~6, p. 066111, 2004.

\bibitem{meyers2003applying}
L.~A. Meyers, M.~Newman, M.~Martin, and S.~Schrag, ``Applying network theory to
  epidemics: control measures for mycoplasma pneumoniae outbreaks,''
  \emph{Emerging infectious diseases}, vol.~9, no.~2, pp. 204--210, 2003.

\bibitem{cauchemez2011role}
S.~Cauchemez, A.~Bhattarai, T.~L. Marchbanks, R.~P. Fagan, S.~Ostroff, N.~M.
  Ferguson, D.~Swerdlow, S.~V. Sodha, M.~E. Moll, F.~J. Angulo \emph{et~al.},
  ``Role of social networks in shaping disease transmission during a community
  outbreak of 2009 h1n1 pandemic influenza,'' \emph{Proceedings of the National
  Academy of Sciences}, vol. 108, no.~7, pp. 2825--2830, 2011.

\bibitem{newman2002spread}
M.~E. Newman, ``Spread of epidemic disease on networks,'' \emph{Physical review
  E}, vol.~66, no.~1, p. 016128, 2002.

\bibitem{meyers2005network}
L.~A. Meyers, B.~Pourbohloul, M.~E. Newman, D.~M. Skowronski, and R.~C.
  Brunham, ``Network theory and sars: predicting outbreak diversity,''
  \emph{Journal of theoretical biology}, vol. 232, no.~1, pp. 71--81, 2005.

\bibitem{salathe2010high}
M.~Salath{\'e}, M.~Kazandjieva, J.~W. Lee, P.~Levis, M.~W. Feldman, and J.~H.
  Jones, ``A high-resolution human contact network for infectious disease
  transmission,'' \emph{Proceedings of the National Academy of Sciences}, vol.
  107, no.~51, pp. 22\,020--22\,025, 2010.

\bibitem{bansal2010dynamic}
S.~Bansal, J.~Read, B.~Pourbohloul, and L.~A. Meyers, ``The dynamic nature of
  contact networks in infectious disease epidemiology,'' \emph{Journal of
  Biological Dynamics}, vol.~4, no.~5, pp. 478--489, 2010.

\bibitem{bettencourt2006power}
L.~M. Bettencourt, A.~Cintr{\'o}n-Arias, D.~I. Kaiser, and
  C.~Castillo-Ch{\'a}vez, ``The power of a good idea: Quantitative modeling of
  the spread of ideas from epidemiological models,'' \emph{Physica A:
  Statistical Mechanics and its Applications}, vol. 364, pp. 513--536, 2006.

\bibitem{bakshy2012role}
E.~Bakshy, I.~Rosenn, C.~Marlow, and L.~Adamic, ``The role of social networks
  in information diffusion,'' in \emph{Proceedings of the 21st international
  conference on World Wide Web}.\hskip 1em plus 0.5em minus 0.4em\relax ACM,
  2012, pp. 519--528.

\bibitem{kempe2003maximizing}
D.~Kempe, J.~Kleinberg, and {\'E}.~Tardos, ``Maximizing the spread of influence
  through a social network,'' in \emph{Proceedings of the ninth ACM SIGKDD
  international conference on Knowledge discovery and data mining}.\hskip 1em
  plus 0.5em minus 0.4em\relax ACM, 2003, pp. 137--146.

\bibitem{zhang2011research}
Z.~Y.-C. L.~Y. Zhang and H.-F. C. H.~X. Fei, ``The research of information
  dissemination model on online social network [j],'' \emph{Acta Physica
  Sinica}, vol.~5, p. 010, 2011.

\bibitem{gruhl2004information}
D.~Gruhl, R.~Guha, D.~Liben-Nowell, and A.~Tomkins, ``Information diffusion
  through blogspace,'' in \emph{Proceedings of the 13th international
  conference on World Wide Web}.\hskip 1em plus 0.5em minus 0.4em\relax ACM,
  2004, pp. 491--501.

\bibitem{saxena2015understanding}
A.~Saxena, S.~Iyengar, and Y.~Gupta, ``Understanding spreading patterns on
  social networks based on network topology,'' \emph{arXiv preprint
  arXiv:1505.00457}, 2015.

\bibitem{content}
J.~Berger and K.~Milkman, ``Social transmission, emotion, and the virality of
  online content,'' \emph{Wharton Research Paper}, 2010.

\bibitem{online}
J.~Berger and K.~L. Milkman, ``What makes online content viral?'' \emph{Journal
  of Marketing Research}, vol.~49, no.~2, pp. 192--205, 2012.

\bibitem{textviral}
M.~Guerini, C.~Strapparava, and G.~{\"O}zbal, ``Exploring text virality in
  social networks.'' in \emph{ICWSM}, 2011.

\bibitem{hashtag}
O.~Tsur and A.~Rappoport, ``What's in a hashtag?: content based prediction of
  the spread of ideas in microblogging communities,'' in \emph{Proceedings of
  the fifth ACM international conference on Web search and data mining}.\hskip
  1em plus 0.5em minus 0.4em\relax ACM, 2012, pp. 643--652.

\bibitem{digg}
K.~Lerman and R.~Ghosh, ``Information contagion: An empirical study of the
  spread of news on digg and twitter social networks.'' \emph{ICWSM}, vol.~10,
  pp. 90--97, 2010.

\bibitem{natures}
L.~Weng, F.~Menczer, and Y.-Y. Ahn, ``Virality prediction and community
  structure in social networks,'' \emph{Scientific reports}, vol.~3, 2013.

\bibitem{weng2014predicting}
------, ``Predicting successful memes using network and community structure,''
  \emph{arXiv preprint arXiv:1403.6199}, 2014.

\bibitem{mff}
M.~Cha, H.~Haddadi, F.~Benevenuto, and P.~K. Gummadi, ``Measuring user
  influence in twitter: The million follower fallacy.'' \emph{ICWSM}, vol.~10,
  pp. 10--17, 2010.

\bibitem{influence}
M.~Kitsak, L.~K. Gallos, S.~Havlin, F.~Liljeros, L.~Muchnik, H.~E. Stanley, and
  H.~A. Makse, ``Identification of influential spreaders in complex networks,''
  \emph{Nature Physics}, vol.~6, no.~11, pp. 888--893, 2010.

\bibitem{alvarez2005k}
J.~I. Alvarez-Hamelin, L.~Dall'Asta, A.~Barrat, and A.~Vespignani, ``k-core
  decomposition of internet graphs: hierarchies, self-similarity and
  measurement biases,'' \emph{arXiv preprint cs/0511007}, 2005.

\end{thebibliography}

\end{document}